\documentstyle[preprint,aps]{revtex}

\begin{document}

\draft

\tighten

\preprint{\vbox{\hfill UCSB--TH--95--34 \\
          \vbox{\hfill chao-dyn/9511001} \\
          \vbox{\hfill November 1995}
          \vbox{\vskip1.0in}
         }}

\title{Thermal fluctuations in quantized chaotic systems}

\author{Mark Srednicki\footnote{E--mail: \tt mark@tpau.physics.ucsb.edu}}

\address{Department of Physics, University of California,
         Santa Barbara, CA 93106
         \\ \vskip0.5in}

\maketitle

\begin{abstract}
\normalsize{
We consider a quantum system with $N$ degrees of freedom which is classically
chaotic.  When $N$ is large, and both $\hbar$ and the quantum energy
uncertainty $\Delta E$ are small, quantum chaos theory can be used to
demonstrate the following results: (1) given a generic observable $A$, the
infinite time average $\overline A$ of the quantum expectation value
$\langle A(t)\rangle$ is independent of all aspects of the initial state
other than the total energy, and equal to an appropriate thermal average
of $A$; (2) the time variations of $\langle A(t)\rangle - \overline A$
are too small to represent thermal fluctuations; (3) however, the time
variations of $\langle A^2(t)\rangle - \langle A(t)\rangle^2$ can be
consistently interpreted as thermal fluctuations, even though these same
time variations would be called quantum fluctuations when $N$ is small.
}
\end{abstract}


In this paper we examine the compatibility of certain results in quantum
chaos theory with standard results in statistical mechanics.  We consider
a bounded, isolated, many-body quantum system whose classical limit is chaotic.
Given an initial state $|\psi(0)\rangle$ and a generic observable $A$,
we ask the following questions.  What is the infinite time average of
$\langle A(t)\rangle \equiv \langle \psi(t)|A|\psi(t)\rangle$?
Is it independent of the initial state $|\psi(0)\rangle$?
If so, is it equal to an appropriate thermal average of $A$?
What are the root-mean-square fluctuations, in time,
of $\langle A(t)\rangle$ about its infinite time average?
Are these fluctuations correctly predicted by statistical mechanics?

We begin by noting that the energy spectrum of a bounded quantum
system is purely discrete; if the system is classically chaotic,
and also has no discrete symmetries, then the energy eigenvalues
$E_\alpha$ are almost always nondegenerate \cite{blec1}.
Since we assume that the system is isolated, its state at time $t$ is
\begin{equation}
|\psi(t)\rangle =
\sum_\alpha C_\alpha\,e^{-iE_\alpha t/\hbar}\,|\alpha\rangle \,,
\label{psit}
\end{equation}
where the $C_\alpha$'s specify the initial state, and we assume the
usual normalization
\begin{equation}
\sum_\alpha |C_\alpha|^2 = 1 \,.
\label{norm}
\end{equation}
The expectation value of an observable $A$ at time $t$ is
\begin{eqnarray}
\langle A(t)\rangle &\equiv& \langle\psi(t)|A|\psi(t)\rangle
\nonumber \\
\noalign{\medskip}
    &=& \sum_{\alpha\beta}C^*_\alpha C^{\phantom{*}}_\beta\,
        e^{i(E_\alpha-E_\beta)t/\hbar} A_{\alpha\beta} \,,
\label{at}
\end{eqnarray}
where
\begin{equation}
A_{\alpha\beta} \equiv \langle\alpha|A|\beta\rangle
\label{matel}
\end{equation}
are the matrix elements of $A$ in the energy eigenstate basis.
The infinite time average of $\langle A(t)\rangle$ is given by
\begin{eqnarray}
\overline{A} &\equiv& \lim_{\tau\to\infty}{1\over\tau}
                      \int_0^\tau dt\;\langle A(t)\rangle
\nonumber \\
\noalign{\medskip}
             &=& \sum_\alpha |C_\alpha|^2 A_{\alpha\alpha} \,.
\label{ita}
\end{eqnarray}
The time averaged fluctuations of $\langle A(t)\rangle$ about $\overline A$
are given by
\begin{eqnarray}
\overline{ \left[\langle A(t)\rangle-\overline{A}\,\right]^2 }
&\equiv& \lim_{\tau\to\infty}{1\over\tau}
         \int_0^\tau dt\,\left[\langle A(t)\rangle-\overline{A}\,\right]^2
\nonumber \\
\noalign{\medskip}
             &=& \sum_{\alpha,\beta\ne\alpha} |C_\alpha|^2\,|C_\beta|^2\,
                                              |A_{\alpha\beta}|^2 \,.
\label{deltaita}
\end{eqnarray}
We now turn to a discussion of what can be inferred about (\ref{ita}) and
(\ref{deltaita}) from quantum chaos theory.

Quantum chaos theory is largely based on semiclassical arguments; to make
use of it, we will have to assume that Planck's constant is ``small.''
This means that there is some dimensionless combination of parameters,
with a single power of Planck's constant in the numerator, which serves as
an expansion parameter for quantities such as $A_{\alpha\beta}$.
The relevant combination of parameters, which we will call $\hbar$,
obviously depends on the system under consideration.  How small $\hbar$
has to be depends on both $A$ and the range of energies which are of interest.
It is particularly difficult to determine the dependence of $\hbar$ on $N$,
the number of degrees of freedom in the system.  This question is
irrelevant when $N$ is small, but crucial when $N$ is large.
We will not discuss this important problem any further here, however;
we will simply assume that the correct expansion parameter $\hbar$,
whatever its dependence on $N$, is sufficiently small.

Given a classically chaotic system with $N$ degrees of freedom, we consider
an observable $A$ which is a smooth function of the classical coordinates and
momenta, and which has no explicit dependence on $\hbar$.  Then quantum chaos
theory predicts that the matrix elements $A_{\alpha\beta}$ are given by
\begin{equation}
A_{\alpha\beta} = {\cal A}(E_\alpha)\delta_{\alpha\beta}
                  + \hbar^{(N-1)/2} R_{\alpha\beta} \,.
\label{semic}
\end{equation}
Here ${\cal A}(E)$ is a smooth function of energy whose leading term in the
$\hbar$ expansion is $O(\hbar^0)$.  The matrix elements $R_{\alpha\beta}$
are also $O(\hbar^0)$ at leading order, and their values are characterized by
a smooth distribution, often assumed to be gaussian.  Eq.~(\ref{semic}) has
not been demonstrated rigorously, but it follows from a variety of different
arguments, including Berry's random-wave conjecture for the energy
eigenfunctions \cite{randwave,fppro,me},
the analogy between quantized chaotic systems and random matrix
theory \cite{randmat}, and the semiclassical periodic orbit expansion,
assuming a certain randomness for the periodic orbits \cite{perorb}.
There is, however, one aspect of (\ref{semic}) which has been proven
rigorously; specifically,
\begin{equation}
\lim_{\alpha\to\infty} A_{\alpha\alpha} =
{\int d^N\!p\,d^N\!q\;\delta(H(p,q)-E_\alpha) A(p,q) \over
 \int d^N\!p\,d^N\!q\;\delta(H(p,q)-E_\alpha)} \,,
\label{limit}
\end{equation}
where $H(p,q)$ is the classical hamiltonian, and $A(p,q)$ is the classical
form of the operator $A$ \cite{shnrl}.  The limit holds for all energy
eigenstates $|\alpha\rangle$ except possibly a subsequence of density zero.
The right-hand side of (\ref{limit}) is the $O(\hbar^0)$ contribution to
${\cal A}(E_\alpha)$.

For later use, we must also examine the matrix elements of $A^2$.
Consider first the diagonal elements
$(A^2)_{\alpha\alpha} = \sum_\beta A_{\alpha\beta}A_{\beta\alpha}$;
using (\ref{semic}) gives
\begin{equation}
(A^2)_{\alpha\alpha} = \biggl[{\cal A}^2(E_\alpha) +
               \hbar^{N-1}\sum_\beta |R_{\alpha\beta}|^2\biggr]
    + \hbar^{(N-1)/2}\,2{\cal A}(E_\alpha) R_{\alpha\alpha} \,.
\label{asq2}
\end{equation}
We have grouped the terms as shown because the second term in square brackets
is actually $O(\hbar^0)$, despite the explicit factor of $\hbar^{N-1}$.
This is because the sum over $\beta$ can be converted to an integral over
the quantum density of states, and the quantum density of states is
$O(\hbar^{-N})$ \cite{blec1}.  One more factor of $\hbar$ then arises
from converting a quantum energy integral into a classical frequency
integral \cite{fppro}.  Thus, the diagonal matrix elements of $A^2$
have the same general structure (\ref{semic}) as the diagonal matrix
elements of $A$; this is of course required for internal consistency,
since there was nothing special about $A$.

Now consider the off-diagonal elements
$(A^2)_{\alpha\gamma} = \sum_\beta A_{\alpha\beta}A_{\beta\gamma}$;
using (\ref{semic}) gives
\begin{equation}
(A^2)_{\alpha\gamma}
= \hbar^{(N-1)/2} [{\cal A}(E_\alpha) +{\cal A}(E_\gamma)] R_{\alpha\gamma}
 + \hbar^{N-1}\sum_\beta R_{\alpha\beta}R_{\beta\gamma}
\label{asq3}
\end{equation}
when $\alpha\ne\gamma$.  This time, however, the sum over $\beta$ in the
last term does not contribute a factor of $\hbar^{-N+1}$,
because $R_{\alpha\beta}R_{\beta\gamma}$ is not positive definite.
Instead, we expect $R_{\alpha\beta}R_{\beta\gamma}$ to have a phase
(or perhaps just a sign) which varies erratically with $\beta$.
This implies that the sum over $\beta$ of $R_{\alpha\beta}R_{\beta\gamma}$
is the same order in $\hbar$ as the square root of the sum over $\beta$ of
$|R_{\alpha\beta}R_{\beta\gamma}|^2$; this latter sum is $O(\hbar^{-N+1})$.
Thus we conclude that, overall, the second term on the right-hand side
of (\ref{asq3}) is $O(\hbar^{(N-1)/2})$, just like the first term,
and just like the off-diagonal matrix elements of $A$.  Again this is required
for the consistency of (\ref{semic}) with the generic character of $A$.

Returning to (\ref{ita}), if we insert (\ref{semic}) we get
\begin{equation}
\overline{A} = \sum_\alpha |C_\alpha|^2 {\cal A}(E_\alpha)
               + O(\hbar^{(N-1)/2}) \,.
\label{ita2}
\end{equation}
We now assume that the expected value of the total energy
\begin{equation}
\langle E\rangle = \sum_\alpha |C_\alpha|^2 E_\alpha
\label{eav}
\end{equation}
has a quantum uncertainty
\begin{equation}
\Delta E = \biggl[\sum_\alpha |C_\alpha|^2\,(E_\alpha-\langle E\rangle)^2
           \biggr]^{1/2}
\label{deltae}
\end{equation}
which is small, in a sense which we will make more precise shortly.
This is a natural assumption if $N$ is large, since states of physical
interest typically have $\Delta E \sim N^{-1/2}\langle E\rangle$.  Note,
however, that in this case the smallness of $\Delta E$ does not imply or
require the smallness of $\hbar$.

Assuming $\Delta E$ is small, we can expand ${\cal A}(E_\alpha)$ about
$\langle E\rangle$ to get
\begin{equation}
{\cal A}(E_\alpha) = {\cal A}(\langle E\rangle) +
(E_\alpha-\langle E\rangle){\cal A}'(\langle E\rangle) +
{\textstyle{1\over2}}(E_\alpha-\langle E\rangle)^2{\cal A}''(\langle E\rangle)
+ \ldots\,.
\label{aexp}
\end{equation}
Substituting this expansion into (\ref{ita2}), we find
\begin{equation}
\overline{A} = {\cal A}(\langle E\rangle) +
{\textstyle{1\over2}}(\Delta E)^2{\cal A}''(\langle E\rangle)
+ O((\Delta E)^3) + O(\hbar^{(N-1)/2}) \,.
\label{ita3}
\end{equation}
Thus, the infinite time average $\overline A$ depends on the expected value
of the total energy $\langle E\rangle$, but is independent of all other
aspects of the initial state, provided that $\hbar$ is small enough to make
the $O(\hbar^{(N-1)/2})$ term negligible, and provided that
\begin{equation}
(\Delta E)^2\left|{{\cal A}''(\langle E\rangle)
                  \over {\cal A}(\langle E\rangle)}\right| \ll 1 \,.
\label{decrit}
\end{equation}
This is the more precise criterion for the smallness of $\Delta E$.

We are now able to make a connection with statistical mechanics.
Mathematically, we can choose the $|C_\alpha|^2$'s to represent a
microcanonical average over an energy range $\Delta E$ centered on
$\langle E\rangle$.  If this $\Delta E$ is chosen to satisfy (\ref{decrit}),
then $\overline A$ is equal to this microcanonical average of $A$.
Alternatively, we can choose the $|C_\alpha|^2$'s to be canonical Boltzmann
weights; the canonical energy dispersion $\Delta E$ is usually smaller than
$\langle E\rangle$ by a factor of $N^{-1/2}$, and therefore the canonical
$\Delta E$ should satisfy (\ref{decrit}) when $N$ is large.  If so, then
$\overline A$ is equal to the canonical thermal average of $A$ at whatever
temperature results in a total energy of $\langle E\rangle$.  Thus, the
function ${\cal A}(E)$ can in principle be calculated, at least up to
corrections which are $O(\hbar^{(N-1)/2})$ and $O(N^{-1})$, by the methods
of canonical statistical mechanics.

Some time ago, Jaynes \cite{jaynes} pointed out that a canonical calculation
of the size of the thermal fluctuations in some observable $A$ must ultimately
be based on demonstrating that $A$ exhibits time variations with the same
root-mean-square amplitude.  To study this issue in the present context,
we first consider the time variations of
$\langle A(t)\rangle - \overline A$.  From (\ref{deltaita}), (\ref{semic}),
and (\ref{norm}), we find
\begin{equation}
\overline{\left[\langle A(t)\rangle-\overline{A}\,\right]^2}=O(\hbar^{N-1}) \,.
\label{deltaita2}
\end{equation}
We see that the fluctuations of $\langle A(t)\rangle$ about $\overline A$ are
small.  This tells us that, whatever the initial value $\langle A(0)\rangle$
happens to be, $\langle A(t)\rangle$ must eventually approach its thermal
average $\overline A$, and then remain near $\overline A$ most of the time.
(We do not, however, learn anything about the time scale of this approach.)
Apparently, under appropriate circumstances quantum chaos can serve as
the dynamical underpinning of certain basic results of statistical mechanics,
an idea which has already appeared in various guises \cite{me,previous}.

On the other hand, (\ref{deltaita2}) is too small to represent the expected
thermal fluctuations of $A$, which are $O(\hbar^0)$.  To find thermal
fluctuations, we must look at the infinite time average of
$\langle A^2(t)\rangle$; this is given by
\begin{eqnarray}
\overline{A^2} &\equiv& \lim_{\tau\to\infty}{1\over\tau}
                        \int_0^\tau dt\;\langle A^2(t)\rangle
\nonumber \\
\noalign{\medskip}
             &=& \sum_\alpha |C_\alpha|^2\,(A^2)_{\alpha\alpha} \,.
\label{asqita}
\end{eqnarray}
We have already seen that the matrix elements of $A^2$ have the same general
structure as the matrix elements of $A$.  Therefore, we can immediately
conclude that $\overline{A^2}$ is equal to a thermal average of $A^2$,
up to corrections which are $O(\hbar^{(N-1)/2})$ and $O(N^{-1})$.

Putting everything together, we conclude that,
up to corrections which are $O(\hbar^{(N-1)/2})$ and $O(N^{-1})$,
the infinite time average of $\langle A^2(t)\rangle - \langle A(t)\rangle^2$
is equal to a thermal average of $(A-\overline{A}\,)^2$.  Thus, variations
with time of $\langle A^2(t)\rangle - \langle A(t)\rangle^2$ can be interpreted
as representing thermal fluctuations.  It is interesting to note that, in a
few-body system, these same time variations would be called quantum
fluctuations.

To summarize, results from quantum chaos theory are compatible with
results from statistical mechanics; quantum chaos theory can even be used
as a basis from which one can demonstrate, e.g., that the quantum expectation
value of an observable must approach its thermal average, at least when the
number of degrees of freedom $N$ is large, the quantum energy uncertainty
$\Delta E$ is small, and the semiclassical expansion parameter $\hbar$ is
small.  Just how small $\hbar$ needs to be is a question to which we hope
to return.  Also, we have seen that the variations
with time of a quantum expectation value are too small to account for the
expected thermal fluctuations; instead, what would be called quantum
fluctuations when $N$ is small have just the right amplitude to be identified
as thermal fluctuations when $N$ is large.

\vskip0.25in

I would like to thank John Preskill for helpful discussions.
This work was supported in part by NSF Grant PHY--91--16964.


\begin{references}

\bibitem{blec1} M. V. Berry,
in {\em Les Houches XXXVI, Chaotic Behavior of Deterministic Systems},
edited by G.~Iooss, R. H. G. Helleman, and R. Stora
(North-Holland, Amsterdam, 1983).

\bibitem{randwave} M. V. Berry, J. Phys. A {\bf 10}, 2083 (1977);
P. Pechukas, Phys. Rev. Lett. {\bf 51}, 943 (1983).

\bibitem{fppro} M. Feingold and A. Peres, Phys. Rev. A {\bf 34}, 591 (1986);
T. Prosen, Ann. Phys. {\bf 235}, 115 (1994).

\bibitem{me} M. Srednicki, Phys. Rev. E {\bf 50}, 888 (1994).

\bibitem{randmat} O. Bohigas, M.-J. Giannoni, and C. Schmit,
Phys. Rev. Lett. {\bf 52}, 1 (1984);
O. Bohigas, in {\em Les Houches LII, Chaos and Quantum Physics},
edited by M.-J. Giannoni, A.~Voros, and J. Zinn-Justin
(North--Holland, Amsterdam, 1991).

\bibitem{perorb} B. Eckhardt and J. Main,
Phys. Rev. Lett. {\bf 75}, 2300 (1995);
B. Eckhardt, S. Fishman, J. Keating, O. Agam, J. Main, and K. M\"uller,
Los Alamos Archive Report No. chao-dyn/9509017, to be published.

\bibitem{shnrl} A. I. Shnirelman, Ups. Mat. Nauk {\bf 29}, 181 (1974);
S. Zelditch, Duke Math J. {\bf 55}, 919 (1987);
Y. Colin de Verdi\`ere, Comm. Math. Phys. {\bf 102}, 497 (1985);
in {\em Les Houches LII, Chaos and Quantum Physics},
edited by M.-J. Giannoni, A.~Voros, and J. Zinn-Justin
(North--Holland, Amsterdam, 1991); B. Helffer, A. Martinez, and D. Robert,
Comm. Math. Phys. {\bf 109}, 313 (1987).

\bibitem{jaynes} E. T. Jaynes, in {\em The Maximum Entropy Formalism},
edited by R. D. Levine and M. Tribus (MIT, Cambridge, 1979).

\bibitem{previous} N. G. van Kampen, in {\em Chaotic Behavior in Quantum
Systems}, edited by G. Casati (Plenum, New York, 1985);
J. M. Deutsch, Phys. Rev. A {\bf 43}, 2046 (1989);
P. Gaspard, in {\em Quantum Chaos -- Quantum Measurement}, edited by
P. Cvitanovi\'c, I. Percival, and A. Wirzba (Kluwer, Dordrecht, 1992).

\end{references}
\end{document}